\documentclass{mem}
\usepackage{natbib}
\usepackage{txfonts}
\usepackage{balance}
\usepackage{graphicx}
\usepackage[a4paper,breaklinks]{hyperref}
\idline{87}{610}
\begin{document}

\newcommand {\Msun} {M$_\odot$}
\newcommand {\Mbss} {M_\mathrm{BSS}}
\newcommand {\Mto} {M_\mathrm{MSTO}}

\title{HST proper motions in Galactic globular clusters}
\subtitle{}

\author{
L.~L.\,Watkins\inst{1,4} \and \\
R.~P.\,van der Marel\inst{1} \and
A.\,Bellini\inst{1} \and
A.~T.\,Baldwin\inst{1,2} \and
P.\,Bianchini\inst{3} \and
J.\,Anderson\inst{1}
}

\institute{
Space Telescope Science Institute, 3700 San Martin Drive, Baltimore MD 21218, USA \and
Dept. of Physics \& Astronomy, Louisiana State Univ., Baton Rouge, LA 70803, USA \and
Max Planck Institute for Astronomy, K\"{o}nigstuhl 17, D-69117 Heidelberg, Germany \and
\email{lwatkins@stsci.edu}
}

\authorrunning{Watkins}

\titlerunning{HST proper motions in Galactic GCs}

\abstract{Proper motions (PMs) are crucial to fully understand the internal dynamics of globular clusters (GCs). To that end, the \textit{Hubble Space Telescope (HST)} Proper Motion (HSTPROMO) collaboration has constructed large, high-quality PM catalogues for 22 Galactic GCs. We highlight some of our exciting recent results: the first directly-measured radial anisotropy profiles for a large sample of GCs; the first dynamical distance and mass-to-light (M/L) ratio estimates for a large sample of GCs; and the first dynamically-determined masses for hundreds of blue-straggler stars (BSSs) across a large GC sample.
\keywords{globular clusters: general -- proper motions -- stars: blue stragglers -- stars: distances -- stars: kinematics and dynamics -- stars: luminosity function, mass function}
}
\maketitle{}

\section{Introduction}

The HSTPROMO collaboration is using PMs to revolutionise our dynamical understanding of many objects in the universe -- including stars in globular and young star clusters; Local Group galaxies, including Andromeda, the Magellanic Clouds and a number of dwarf spheroidals; and even AGN black hole jets -- thanks to the exquisite astrometric precision of \textit{HST} \citep{vandermarel2014}.\footnote{\href{http://www.stsci.edu/~marel/hstpromo.html}{http://www.stsci.edu/$\sim$marel/hstpromo.html}}

As part of this ongoing work, \citet{bellini2014} recently presented a set of internal PM catalogues for 22 Galactic GCs, measured using archival data from \textit{HST}. In \citet{watkins2015a}, \citet{watkins2015b}, and \citet{baldwin2016}, we used these catalogues to study 3 different aspects of the GC sample: 1) velocity anisotropy profiles; 2) dynamical distances and M/Ls; and 3) masses of their BSS populations. Here we briefly highlight the results from each study.

\section{Velocity anisotropy}

Dynamical mass estimates are degenerate with anisotropy, so understanding the anisotropy in a stellar system is crucial to successful mass determination.

In \citet{watkins2015a}, we began by making a series of cuts to select high-quality samples of bright stars. By restricting the magnitude range of the samples to only those stars brighter than 1~mag below the main-sequence turn off (MSTO), we limited the stellar-mass range in each sample, and so could neglect the effect of stellar mass on the kinematics and consider only the spatial changes. The quality cuts were made to eliminate stars for which the PMs were poorly measured or for which the uncertainties had been underestimated as such stars can introduce biases into kinematic analyses. We then constructed binned velocity dispersion and anisotropy profiles for each GC.

\begin{figure}
    \centering
    \includegraphics[width=\linewidth]{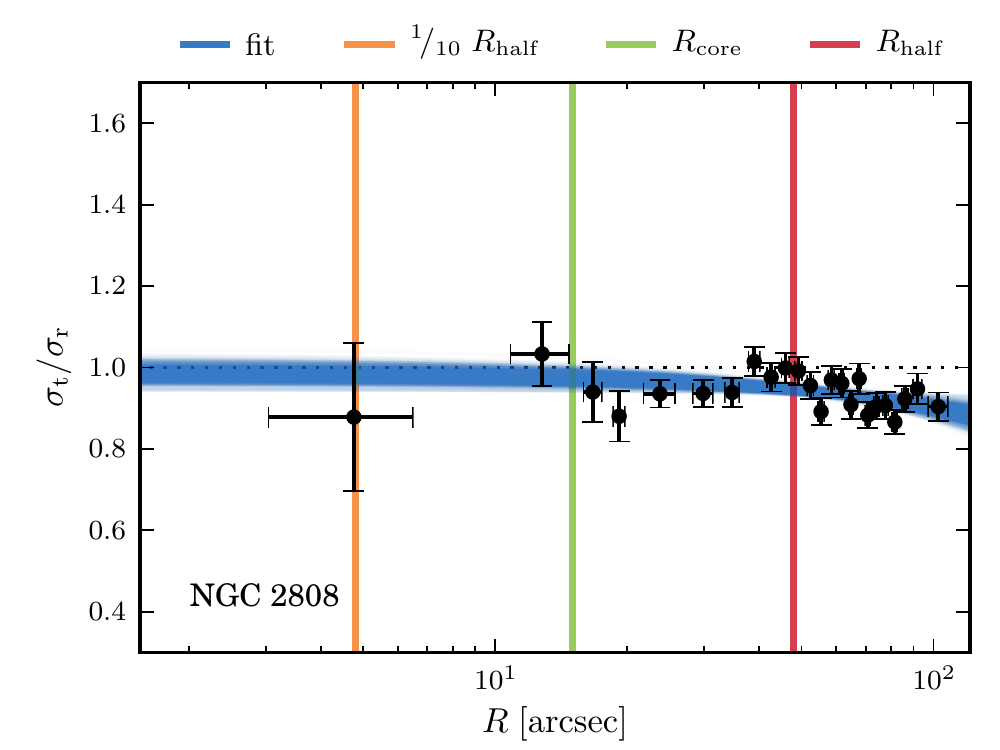}
    \caption{\footnotesize Velocity anisotropy as a function of projected distance from the cluster centre for NGC\,2808. The black points show the binned velocity anisotropy profile and the blue lines show a simple fit. The red (green, orange) line marks the half-light (core, one-tenth of the half-light) radius.}
    \label{2808_aniso}
\end{figure}

\begin{figure}
    \centering
    \includegraphics[width=\linewidth]{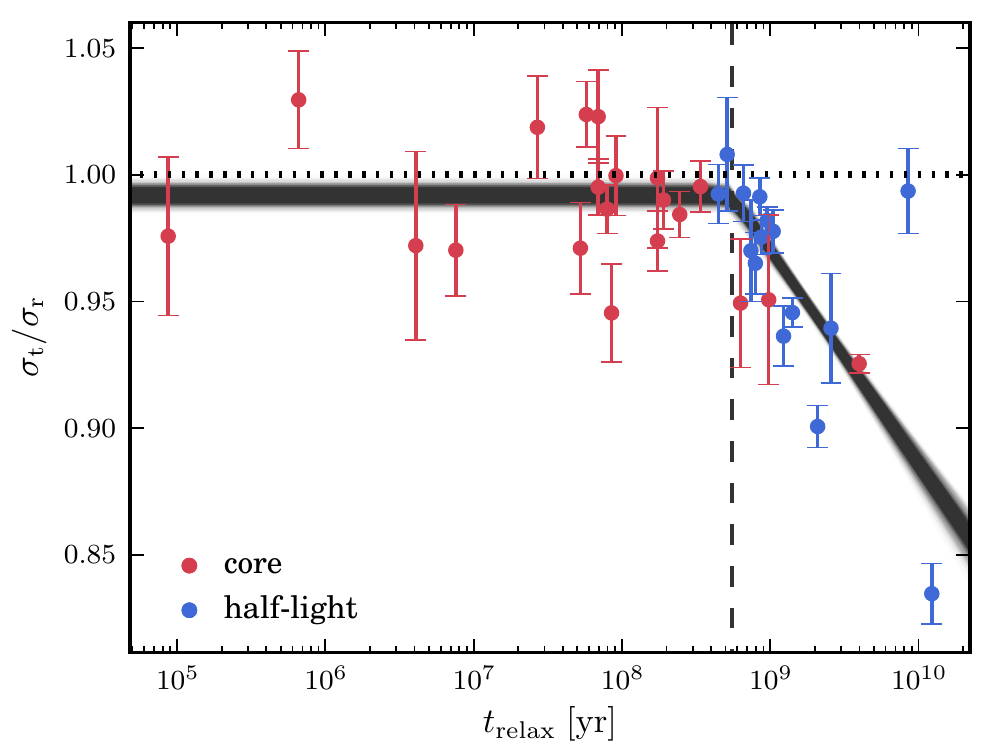}
    \caption{\footnotesize Velocity anisotropy as a function of relaxation time for all GCs in our sample. The red (blue) points show the values estimated at the core (half-light) radius. The GCs are isotropic in regions with relaxation times shorter than a characteristic time (dashed line) and then become increasingly radially anisotropic with increasing relaxation time.}
    \label{aniso_trelax}
\end{figure}

Figure~\ref{2808_aniso} shows the binned anisotropy profile for NGC\,2808 (black points). This GC is isotropic at its centre and becomes mildly radially anisotropic with increasing distance from the centre. This trend is typical for all GCs in our sample; to quantify this, we used the fits (blue lines) to estimate the anisotropy at the core and half-light radii (green and red lines) and compared these values to estimates of the relaxation times at these radii \citep[][2010 edition]{harris1996}. Figure~\ref{aniso_trelax} shows the results of this comparison. Nearly all GCs appear to be isotropic out to their core radii; thereafter, some remain isotropic out to their half-light radii, while others become mildly radially anisotropic, with the degree of anisotropy increasing with relaxation time. The black lines show a fit to the data with a break between the isotropic and anisotropic regions at the characteristic time marked by the dashed line.

This analysis offers a way to estimate the vital anisotropy of a GC using its relaxation time, when no PM data is available.

\section{Dynamical distances and mass-to-light ratios}

GC distances are typically estimated using photometric methods that compare the apparent and absolute magnitudes of stars for which the absolute magnitudes are known or may be inferred, such as RR Lyrae stars. M/Ls are typically inferred from via stellar population synthesis (SPS) modelling. However, both distances and M/Ls can be estimated using dynamical modelling when both PM and line-of-sight (LOS) velocity data exist. The photometric and dynamical methods use very different types of data to constrain the same fundamental properties, so their comparison can serve as a crucial test of both methods.

In \citet{watkins2015b}, we used cleaned samples of bright stars to construct PM velocity dispersion profiles and then compared these against LOS velocity dispersion profiles from the literature. This was only possible for 15 of the 22 GCs, the remaining GCs had insufficient (or even no) LOS data available. From this analysis, we estimated dynamical distances and M/Ls for each GC, which we compared against photometric distances from \citet[][2010 edition]{harris1996} and SPS M/Ls from \citet{mclaughlin2005}.

\begin{figure}
    \centering
    \includegraphics[width=\linewidth]{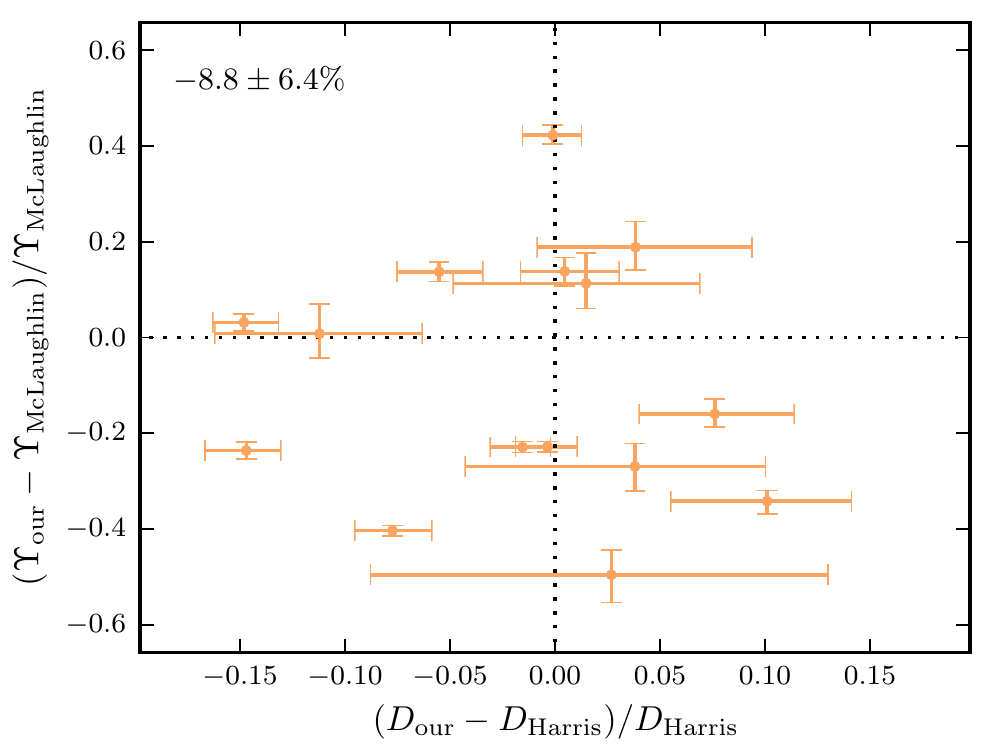}
    \caption{\footnotesize Fractional difference between the dynamical and photometric distances versus fractional difference between the dynamical and SPS M/Ls. Both the distances and the M/Ls show pleasing agreement, highlighting the robustness of both dynamical and photometric methods.}
    \label{distml_comp}
\end{figure}

Figure~\ref{distml_comp} shows the fractional difference in the dynamical and photometric distances versus the fractional difference in the dynamical and SPS M/Ls. The mean difference in the distances was just $-1.7 \pm 1.9 \%$, indicating excellent agreement and highlighting the robustness of both methods. The mean difference in the M/Ls was $-8.8 \pm 6.4 \%$, showing slightly more scatter but still consistent within 1.3$\sigma$.

\begin{figure}
    \centering
    \includegraphics[width=\linewidth]{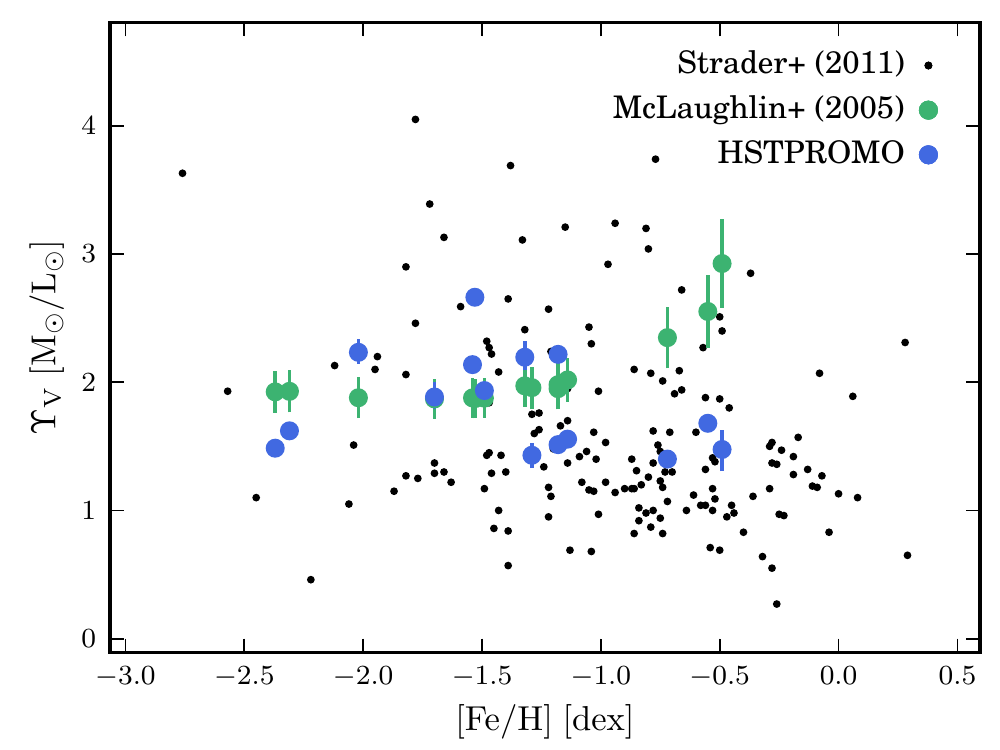}
    \caption{\footnotesize M/L estimates as a function of metallicity. Blue points show our dynamical M/Ls and green points show SPS M/Ls. The SPS models predict an upturn in M/L for metal-rich GCs, whereas our dynamical M/Ls predict a downturn. This behaviour is consistent with a study of 200 M31 GCs by \citet{strader2011} (black points).}
    \label{ml_feh}
\end{figure}

Figure~\ref{ml_feh} shows the M/Ls as a function of GC metallicity \citet[][2010 edition]{harris1996}. Our dynamical M/Ls are shown in blue and the SPS M/Ls are shown in green. We see that the dynamical and SPS M/Ls are consistent for the metal-poor GCs ([Fe/H]$<-1$~dex), but that they diverge for the metal-rich GCs: the SPS M/Ls increase with increasing metallicity, whereas the dynamical M/Ls decrease. This is consistent with the behaviour noted in a study of 200 M31 GCs by \citet{strader2011} (black points) and has been attributed to the effects of mass segregation \citep{shanahan2015}.

\section{Blue-straggler kinematics and dynamical mass estimates}

Frequent two-body stellar interactions in GCs allow the stars to exchange energy; over time, the stars move towards a state of energy equipartition, where they all have the same energy. As a result, high mass stars tend to move more slowly than low mass stars; this is true even if the GC is only in partial equipartition. This effect can be expressed as $\sigma \propto M^{-\eta}$ (1), where $\sigma$ is the velocity dispersion of a stellar population of mass $M$, and $0 \le \eta \le 0.5$ quantifies the degree of equipartition in the GC.

BSSs are an apparent extension of the main-sequence in a GC, bluer and brighter than the MSTO. Most stars brighter than the MSTO in a GC are evolved stars, with approximately equal masses as the latter stages of stellar evolution are so fast. However, BSSs are believed to have formed via mass-transfer or stellar collisions within a binary system, thus making them a more massive population. So, as a result of equipartition in a GC, we expect them to be moving more slowly.

\begin{figure}
    \centering
    \includegraphics[width=\linewidth]{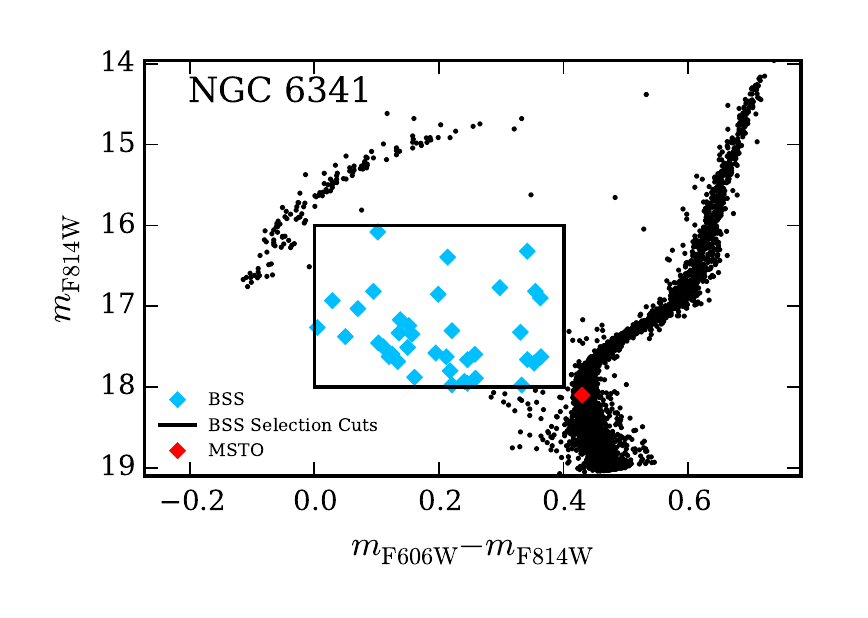}
    \caption{\footnotesize CMD for NGC\,6341. The BSSs are shown as blue diamonds and the box shows the cuts made to select them. The black points show the evolved-star sample, and the red diamond marks the MSTO.}
    \label{bss_cmd}
\end{figure}

\begin{figure}
    \centering
    \includegraphics[width=\linewidth]{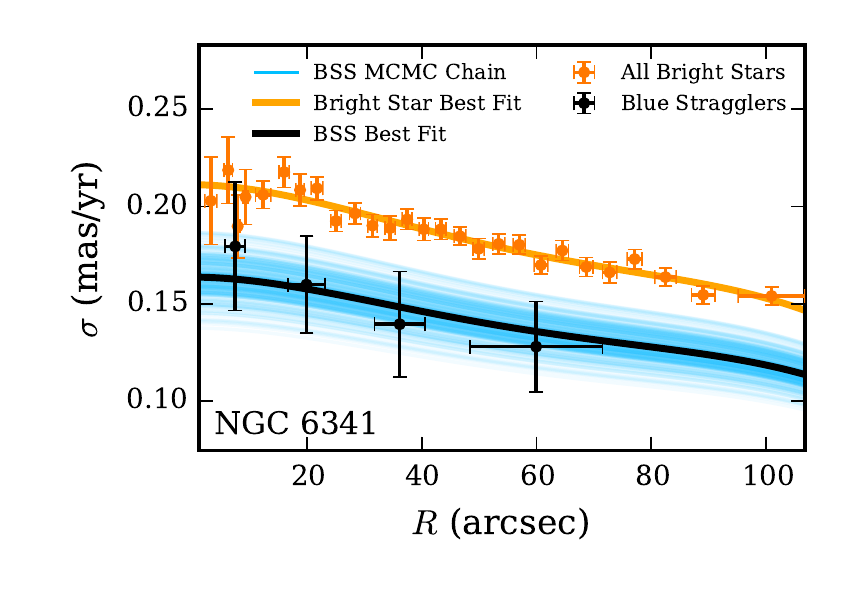}
    \caption{\footnotesize Binned PM dispersion profiles for NGC\,6341. The orange points show the profile for the evolved-stars and the orange line shows a fit to the points. The black points show the profile for the BSSs; the black line is a best fit to the BSS profile that is assumed to be a scaled version of the orange line, the blue lines show the scatter in the fit.}
    \label{bss_disp}
\end{figure}

In \citet{baldwin2016}, we used a series of colour and magnitude cuts to select samples of BSSs in 19 of our 22 GCs, finding 598 BSSs in total. We then calculated binned velocity dispersion profiles for the BSS subsamples and for the evolved stars. Figure~\ref{bss_cmd} shows the colour-magnitude diagram (CMD) for NGC\,6341; the box shows the cuts used to select the BSSs (blue diamonds). The black points show the evolved stars and the red diamond marks the MSTO. Figure~\ref{bss_disp} shows the dispersion profiles for the BSSs (black) and the evolved stars (orange) in NGC\,6341.

On average, we found that the BSS dispersions were lower than the evolved-star dispersions, indicating that the BSSs are indeed more massive. Furthermore, by estimating the degree of equipartition in each GC from the series of N-body simulations presented in \citet{bianchini2016b}, we were able to use equation (1) to estimate the average BSS mass $\Mbss$ in each GC as a function of the MSTO mass $\Mto$. Then by estimating the MSTO mass in each GC, we were thus able to estimate the mass of each BSS population. We found an mass ratio of $\Mbss/\Mto = 1.50 \pm 0.14$ and an average mass $\Mbss = 1.22 \pm 0.12$~\Msun, in good agreement with previous BSS mass estimates.

\section{Conclusions}

PMs are crucial to fully understand the internal dynamics of GCs. To that end, the HSTPROMO collaboration has constructed large, high-quality PM catalogues for 22 Galactic GCs. We highlighted some of our exciting recent results: the first directly-measured radial anisotropy profiles for a large sample of GCs; the first dynamical distance and M/L estimates for a large sample of GCs; and the first dynamically-determined masses for hundreds of BSSs across a large GC sample.

\begin{acknowledgements}
Support for this work was provided by grants for \textit{HST} programs AR-12845 (PI: Bellini) and AR-12648 (PI: van~der~Marel), provided by the Space Telescope Science Institute, which is operated by AURA, Inc., under NASA contract NAS 5-26555.
\end{acknowledgements}

\bibliographystyle{aasjournal}
\bibliography{refs}

\begin{thebibliography}{10}
\expandafter\ifx\csname natexlab\endcsname\relax\def\natexlab#1{#1}\fi

\bibitem[{{Baldwin} {et~al.}(2016){Baldwin}, {Watkins}, {van der Marel},
  {Bianchini}, {Bellini}, \& {Anderson}}]{baldwin2016}
{Baldwin}, A.~T., {Watkins}, L.~L., {van der Marel}, R.~P., {et~al.} 2016,
  ArXiv e-prints

\bibitem[{{Bellini} {et~al.}(2014){Bellini}, {Anderson}, {van der Marel},
  {Watkins}, {King}, {Bianchini}, {Chanam{\'e}}, {Chandar}, {Cool}, {Ferraro},
  {Ford}, \& {Massari}}]{bellini2014}
{Bellini}, A., {Anderson}, J., {van der Marel}, R.~P., {et~al.} 2014, \apj,
  797, 115

\bibitem[{{Bianchini} {et~al.}(2016){Bianchini}, {van de Ven}, {Norris},
  {Schinnerer}, \& {Varri}}]{bianchini2016b}
{Bianchini}, P., {van de Ven}, G., {Norris}, M.~A., {Schinnerer}, E., \&
  {Varri}, A.~L. 2016, \mnras, 458, 3644

\bibitem[{{Harris}(1996)}]{harris1996}
{Harris}, W.~E. 1996, \aj, 112, 1487

\bibitem[{{McLaughlin} \& {van der Marel}(2005)}]{mclaughlin2005}
{McLaughlin}, D.~E., \& {van der Marel}, R.~P. 2005, \apjs, 161, 304

\bibitem[{{Shanahan} \& {Gieles}(2015)}]{shanahan2015}
{Shanahan}, R.~L., \& {Gieles}, M. 2015, \mnras, 448, L94

\bibitem[{{Strader} {et~al.}(2011){Strader}, {Caldwell}, \&
  {Seth}}]{strader2011}
{Strader}, J., {Caldwell}, N., \& {Seth}, A.~C. 2011, \aj, 142, 8

\bibitem[{{van der Marel} {et~al.}(2014){van der Marel}, {Anderson}, {Bellini},
  {Besla}, {Bianchini}, {Boylan-Kolchin}, {Chaname}, {Deason}, {Do},
  {Guhathakurta}, {Kallivayalil}, {Lennon}, {Massari}, {Meyer}, {Platais},
  {Sabbi}, {Sohn}, {Soto}, {Trenti}, \& {Watkins}}]{vandermarel2014}
{van der Marel}, R.~P., {Anderson}, J., {Bellini}, A., {et~al.} 2014, in
  Astronomical Society of the Pacific Conference Series, Vol. 480, , 43

\bibitem[{{Watkins} {et~al.}(2015{\natexlab{a}}){Watkins}, {van der Marel},
  {Bellini}, \& {Anderson}}]{watkins2015a}
{Watkins}, L.~L., {van der Marel}, R.~P., {Bellini}, A., \& {Anderson}, J.
  2015{\natexlab{a}}, \apj, 803, 29

\bibitem[{{Watkins} {et~al.}(2015{\natexlab{b}}){Watkins}, {van der Marel},
  {Bellini}, \& {Anderson}}]{watkins2015b}
---. 2015{\natexlab{b}}, \apj, 812, 149

\end{thebibliography}

\end{document}